\documentclass[journal]{IEEEtran}
\usepackage{amsmath,amsfonts}
\usepackage{amssymb}
\usepackage{algorithmic}
\usepackage{algorithm}
\usepackage{array}
\usepackage[caption=false,font=normalsize,labelfont=sf,textfont=sf]{subfig}
\usepackage{textcomp}
\usepackage{stfloats}
\usepackage{url}
\usepackage{verbatim}
\usepackage{graphicx}
\usepackage{cite}
\usepackage{wasysym}
\usepackage[symbol]{footmisc}
\usepackage{xcolor}
\usepackage{color,soul}
\usepackage{siunitx}
\usepackage{orcidlink}
\usepackage{upgreek}
\usepackage{ulem}
\usepackage[absolute]{textpos}
\textblockorigin{0in}{0in} 
\begin{document}
\title{\fontsize{24}{28}\selectfont Microwave Vortex Motion Characterization of Nb$_3$Sn Coatings for Applications in High Magnetic Fields}
\author{
Pablo~Vidal~García$^{1}$~\orcidlink{0000-0003-4847-1234},~\IEEEmembership{Member,~IEEE}, Andrea~Alimenti$^{1,2}$~\orcidlink{0000-0002-4459-6147},~\IEEEmembership{Member,~IEEE}, 
Dorothea~Fonnesu$^{3}$~\orcidlink{0000-0002-5430-948X},~\IEEEmembership{Member,~IEEE},
Davide~Ford$^{3}$~\orcidlink{0009-0006-8144-5360}, 
Alessandro~Magalotti$^{1,2}$~\orcidlink{0009-0004-3352-1977},~\IEEEmembership{Student Member,~IEEE}, Giovanni~Marconato$^{4}$~\orcidlink{0009-0003-1511-8765},
Cristian~Pira$^{3}$~\orcidlink{0000-0002-5893-1567}, 
Sam~Posen$^{5}$~\orcidlink{0000-0002-6499-306X},
Enrico~Silva$^{1,2}$~\orcidlink{0000-0001-8633-4295},~\IEEEmembership{Senior~Member,~IEEE}, 
Kostiantyn~Torokhtii$^{1}$~\orcidlink{0000-0002-3420-3864},~\IEEEmembership{Member,~IEEE}, 
Nicola~Pompeo$^{1,2}$~\orcidlink{0000-0003-4847-1234},~\IEEEmembership{Senior~Member,~IEEE}
\thanks{Received 17 October 2025; revised 26 January 2026; accepted 29 January 2026. Date of publication 2 February 2026; date of current version 13 February 2026. This work was supported in part by INFN-CSN5 Projects “SAMARA” and “SuperMAD,” in part by the U.S. Department of Energy, Office of Science, National Quantum Information Science Research Centers, Superconducting Quantum Materials and Systems Center (SQMS), under Contract DE-AC0207CH11359, and in part by FermiForward Discovery Group, LLC under Contract 89243024CSC000002 with the U.S. Department of Energy, Office of Science, Office of High Energy Physics. (Corresponding author: Pablo Vidal García.)}
\thanks{Pablo Vidal García and Kostiantyn Torokhtii are with the Department of Industrial, Electronic and Mechanical Engineering, Roma Tre University, 00146 Rome, Italy (e-mail: pablo.vidalgarcia@uniroma3.it).}
\thanks{Andrea Alimenti, Alessandro Magalotti, Enrico Silva, and Nicola Pompeo are with the Department of Industrial, Electronic and Mechanical Engineering, Roma Tre University, 00146 Rome, Italy, and also with INFN, Sezione Roma Tre, 00146 Rome, Italy (e-mail: nicola.pompeo@uniroma3.it).}
\thanks{Dorothea Fonnesu, Davide Ford, and Cristian Pira are with INFN, Laboratori Nazionali di Legnaro, 35020 Legnaro, Italy.}
\thanks{Giovanni Marconato is with the University of Hamburg, 22761 Hamburg, Germany.}
\thanks{Sam Posen is with Fermi National Accelerator Laboratory, Batavia, IL 60510 USA.}
\thanks{Color versions of one or more figures in this article are available at
https://doi.org/10.1109/TASC.2026.3660608.}
\thanks{Digital Object Identifier 10.1109/TASC.2026.3660608}
}
%
\markboth{IEEE TRANSACTIONS ON APPLIED SUPERCONDUCTIVITY, VOL. 36, NO. 5, AUGUST 2026}{}
\maketitle
\begin{textblock}{7}(0.75,10.5) 
\noindent\centering
\copyright~2026 The Authors. This work is licensed under a Creative Commons Attribution 4.0 License. For more information, see https://creativecommons.org/licenses/by/4.0/
\end{textblock}
\begin{abstract}
In this work, microwave measurements carried out in dielectric-loaded resonators exposed to high magnetic fields are exploited to yield the surface impedance of Nb$_3$Sn superconducting coatings deposited via two different techniques: vapor tin diffusion, and DC magnetron sputtering. The obtained data lead to qualitative interpretations on both the Nb$_3$Sn superconducting properties, and vortex-dynamics and pinning, of each coating separately, as well as simple distinctive features when comparing those. When examining the respective surface impedances at varying field, it is expected that the studied films perform at substantially diverse magnitudes of flux-flow resistivity, but also in well-differentiated pinning regimes, yet the obtained surface resistances of both samples are comparable, thus demonstrating that there is room for film optimization at the expense of certain compromise between the parameters involved.
\end{abstract}

\section{Introduction}
\IEEEPARstart{T}{he} use of the vast majority of \textit{Superconducting {RadioFrequency}} (SRF) cavities was traditionally dominated by large-scale experiments for acceleration purposes, where significant power consumption may be saved in the long runs, despite the noticeable cryogenic cost that SRF unavoidably brings about~\cite{Padamsee1998,Padamsee2009}. For the  sustainability of future particle accelerators, replacing traditionally-used Nb with any of the available superconducting compounds is already among the plans of the SRF community~\cite{Venturini2023,Miyazaki2024}, as long as the incessant R\&D campaigns prove it feasible. In this sense, Nb$_3$Sn has reached special attention~\cite{Posen2015,Posen2021}, since it presents one of the highest critical temperatures $T_c$ among the Nb intermetallic compounds~\cite{Dew-Hughes1975}, leading to the enlarged energy gap $\Delta(0) \propto T_c$ with respect to Nb. As the gap (and so $T_c$) roughly rules over the temperature dependence of the surface resistance ${R_s \sim \textrm{exp}[-T_c/T]}$~\cite{Turneaure1991} --which is expected to dominate the SRF response in the Meissner state-- Nb$_3$Sn accelerating cavities could be operated in continuous wave mode at the same cryostat temperature \emph{yet} consuming significantly less power. Thus, great current efforts are being taken in achieving successful deposition of Nb$_3$Sn, using different techniques, recipes and substrates \cite{Fonnesu2026,Posen2017}.

In addition to accelerating cavities, several different applications of SRF have emerged in the recent years, such diverse as, for example, beam screens~\cite{Krkotic2022}, or superconducting qubits used in quantum computers~\cite{Blais2004} and quantum sensors~\cite{Degen2017}. In addition, SRF has met further applications in lab-based New Physics experiments, in cavity haloscopes for (galactic halo ``dark matter") axion-to-photon conversion in presence of high magnetic fields~\cite{Alesini2019}, and between cavities in ``light-shining-through-wall" experiments for hidden (``dark") photon-to-photon direct conversion~\cite{Romanenko2023}. Moreover, the uses of those emerging SRF technologies might widen depending on the working conditions, \textit{e.g.}, high-temperature superconductors projected as beam screens could work in relatively small linear accelerators~\cite{Calatroni2025}; and not even the combined use should be disregarded to possibly achieve better performances, \textit{e.g.}, with qubits coupled to cavities operating at sensitivities below the standard quantum limit~\cite{Dixit2021}.

Keeping in mind the considerations above, one might wonder, for instance, if those successful Nb$_3$Sn depositions in vortex-free accelerating cavities could work in haloscopes operated at high magnetic fields --and if so, up to which extent. Answering this non-trivial question is actually the purpose of several current works (also besides Nb$_3$Sn) and settled collaborations~\cite{SRF}. On this point, at the very first stages of relative research, it has been proposed to use measurements on the surface impedance $Z_s$ of small samples of those coatings as deposited, to ultimately compute the haloscope cavity quality factor $Q_0$ that limits the main figures of merit~\cite{Sikivie1983,Sikivie2021}.
The present work aims at giving preliminary insights into the performance of available coatings at microwave frequencies and under high magnetic fields, which is fundamental for future estimations on $Q_0$ of haloscopes. Specifically, two samples regarding two diverse Nb$_3$Sn deposition techniques:
\begin{enumerate}
    \item[($\#1$)] vapor tin diffusion through a high quality bulk Nb substrate~\cite{Posen2017} (named throughout as VTD), whose Nb$_3$Sn layer thickness $d\approxeq[2-3]\,\upmu\textrm{m}$ (cf.~\cite{Lee2020}, where the STEM cross-sectional picture of the VTD coating is reported), and
    \item[($\#2$)] DC magnetron sputtering on the bulk Cu with a thick ($36\,\upmu\textrm{m}$) intermediate Nb buffer layer to accommodate the film~\cite{Fonnesu2026} (named throughout as DCMS), whose nominal thickness $d\approx7.5\,\upmu\textrm{m}$ controlled by deposition time; 
\end{enumerate}
are evaluated separately. By studying the field dependence of the $Z_s$ excess (from the Meissner state) with the applied magnetic field ($H$) ${\Delta Z_s=Z_s(H;T)-Z_s(0;T)}$, some qualitative aspects on the vortex dynamics may be deduced.

A brief theoretical background is presented in Section~\ref{sec:theory}. In Section~\ref{sec:method}, the measurement methods are shortly explained. Section~\ref{sec:results} is reserved for the results and discussions therein. Finally, in Section~\ref{sec:conclusion}, some conclusions are highlighted, and related prospects on future work are stated.

\section{Theoretical background}
\label{sec:theory}
\begin{figure}[!t]
\centering
\vspace{-0.050in}
\includegraphics[width=0.95\columnwidth]{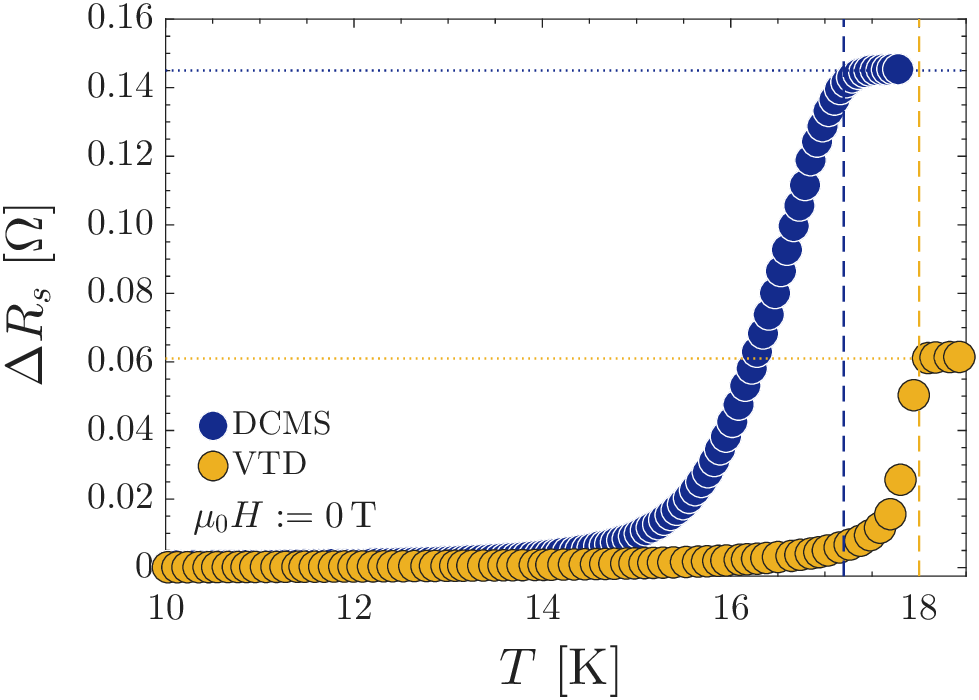}
\makebox[0pt][r]{
    \raisebox{8.7em}{%
      \fbox{\includegraphics[width=.24\linewidth,trim={60cm 22cm 32cm 32cm},clip]{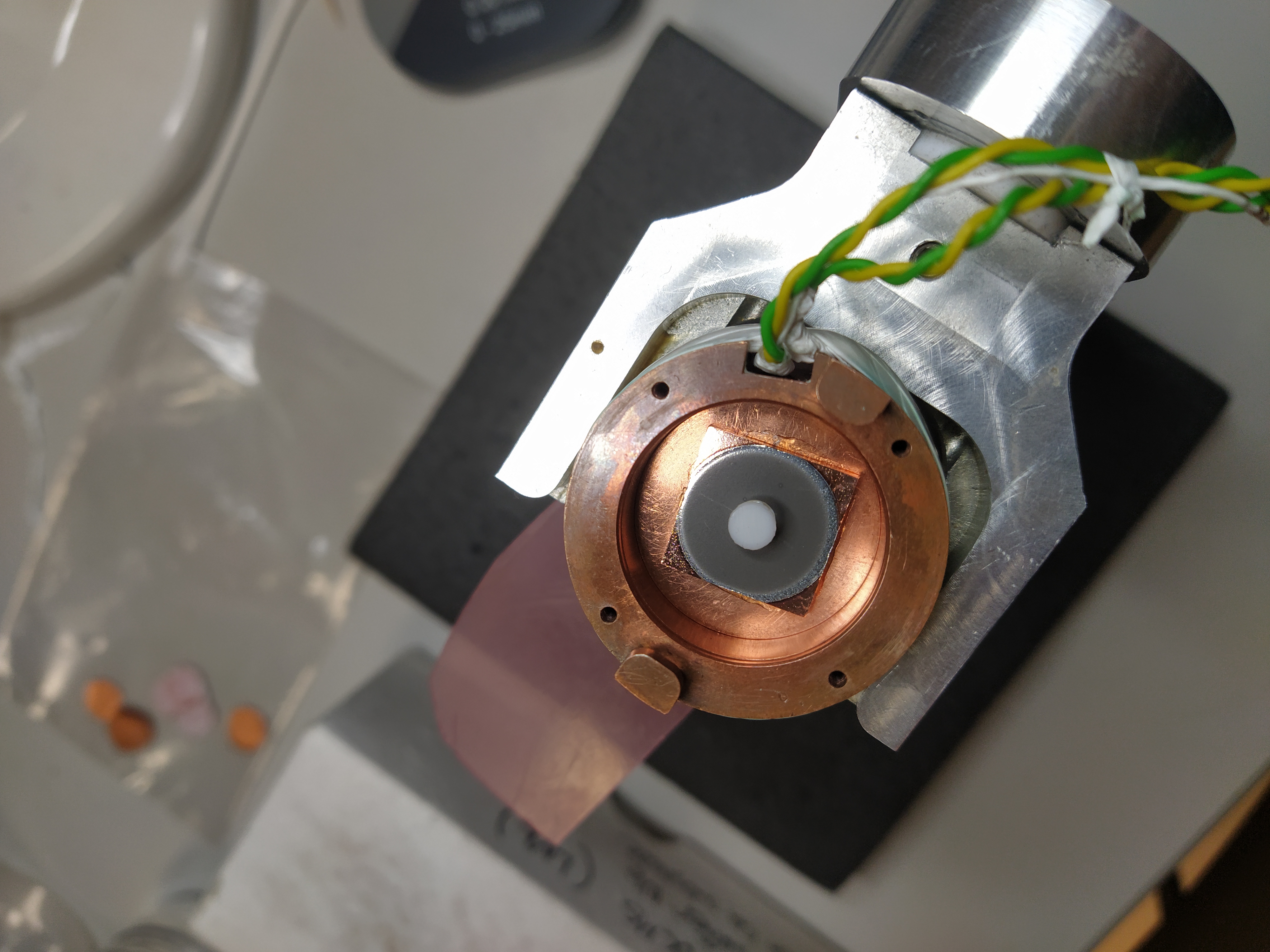}}
    }\hspace*{12.8em}%
    }
\caption{Surface resistance variation of VTD and DCMS samples as a function of temperature ($\Delta R_s(T)$), at zero field. The vertical dashed lines indicate $T_c$, whereas the horizontal dotted lines mark the normal state surface resistance ($R_s\simeq\Delta R_s$). In the inset, the top view of the open DR ($\diameter25\,\textrm{mm}$) with the DCMS sample mounted within is shown. The small dielectric rutile rests on top, which is covered with a thin PTFE cap used as spacer to adjust the coupling. The magnetic field is applied perpendicular to the base of the resonator and sample. The same mounting is used for the VTD sample.}
\label{fig:DR_T-sw}
\end{figure}

The microwave response of type-II bulk superconductors with Ginzburg-Landau parameter ${\kappa\gg1}$ like Nb$_3$Sn is expected to be local~\cite{Hein1999}, thus
\begin{equation}
    Z_s = R_s+\mathrm{i}X_s=\sqrt{{\rm{i}} 2\pi\nu \mu_0 \tilde{\rho}}\textrm{,}
    \label{eq:Zs}
\end{equation}
where $\nu$ is the frequency of the impinging electromagnetic (e.m.) radiation, $R_s$ and $X_s$ are the surface resistance and reactance, respectively, and
\begin{equation}
    \tilde{\rho}=\frac{1-{\rm{i}}\sigma_2\tilde{\rho}_{vm}}{\sigma_1-{\rm{i}}\sigma_2} \simeq \tilde{\rho}_{vm} + {\rm{i}}\lambda^2 2\pi\nu \mu_0
    \label{eq:rho}
\end{equation}
is the complex resistivity accounting for the mean field arising from the motion of each vortex-line through the complex vortex resistivity $\tilde{\rho}_{vm}$~\cite{CoffeyClem1991}, with the induced electric field pointing parallel to the two-fluid ({$\sigma_1-{\rm{i}}\sigma_2$}, where $\sigma_1$ and $\sigma_2$ represent the conductivities associated with quasiparticles and paired electrons in the superconductor, respectively) transport currents~\cite{London1934,Gorter1955,Bardeen1958} induced by e.m. radiation. In the microwave range, at fields far below the upper critical field line, $B_{c2}(T)$, ${\sigma_2\gg\sigma_1}$, therefore the latter approximation in Eq.~\eqref{eq:rho} holds true, in which $\lambda$ is the London penetration length.
For what concerns $\tilde{\rho}_{vm}$, several expressions have been worked out depending on particular features of the superconductor and measurement conditions, yet all of them can be framed within a more general formulation~\cite{Pompeo2008}. In this work we focus on results at low $T$, where flux-creep phenomena can be neglected and the simplified Gittleman-Rosemblum model~\cite{GittlemanRosenblum1966} can be adopted:
\begin{equation}
    \tilde{\rho}_{vm} = \rho_{\textrm{ff}}\frac{{\rm{i}}\nu/\nu_p}{1+{\rm{i}}\nu/\nu_p}
    \textrm{;}
\end{equation}
since it captures the two main contributions to the vortex-motion resistivity, \textit{i.e.},
\begin{enumerate}
    \item[(i)] the scattering of normal electrons at the vortex cores in the flux flow resistivity $\rho_{\textrm{ff}} \propto \rho_n$ ($\rho_n$ is the normal state resistivity), which sets the limit of $\tilde{\rho}_{vm}$ in the absence of pinning; and
    \item[(ii)] the pinning forces acting on vortices in the pinning frequency $\nu_p \propto k_p $ ($k_p$ is the pinning constant that linearizes the mean elastic recall force against pinning energy that traps the vortex at the equilibrium position), which, in broad terms, splits $\nu$ into the pinning and free flow bands, characterized by low vs. high dissipation, respectively;
\end{enumerate}
\begin{figure}[!t]
\centering
\includegraphics[width=1.0\columnwidth]{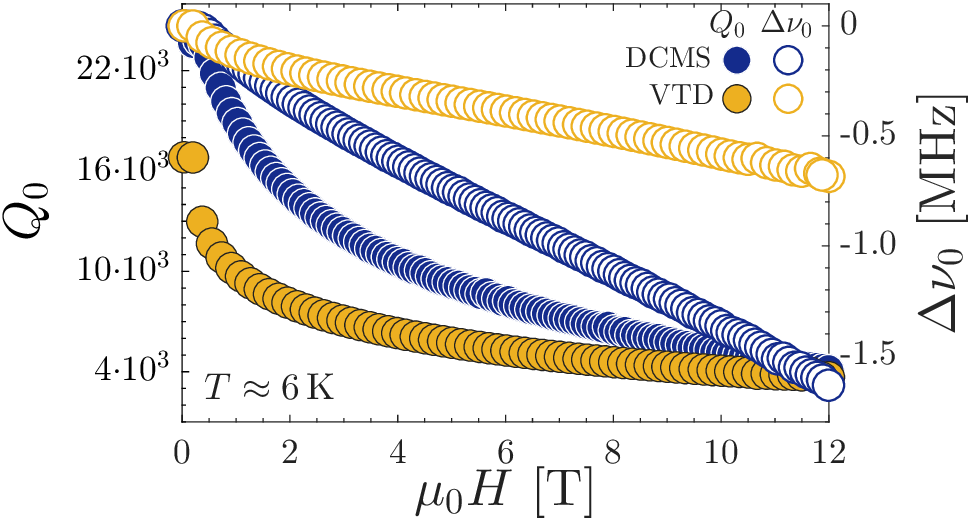}
\caption{$Q_0$ and $\Delta\nu_0$ concerning the measurements of the VTD and DCMS samples as a function of the applied magnetic field, at $\nu_{0}\approx8.50\,\textrm{GHz}$ and $\nu_{0}\approx8.25\,\textrm{GHz}$, respectively, both measured at $T\approx6\,\textrm{K}$.}
\label{fig:Q0Df0}
\end{figure}
while disregarding the vortex inertia, flux creep, and second order Hall effect, which presumably results adequate for low-temperature superconductors tested at microwaves, as stated in previous studies done on Nb$_3$Sn~\cite{Alimenti2020}. In addition, since the e.m. induced currents in the sample are perpendicular to the applied magnetic field in the measurement setup (see the inset of Fig.~\ref{fig:DR_T-sw}), the force exerted on the vortices to probe $\tilde{\rho}_{vm}$ is maximum, a situation which is substantially different in the haloscope cavity~\cite{Sikivie1983,Sikivie2021}, so that one has to account for the relative orientation and possible tilting of fluxons to compute $Z_s$ pointwise on the cavity surface~\cite{Alimenti2022}. Nevertheless, the latter consideration does not change the qualitative discussion on the results presented below.

To conclude this section, it is pointed out that $Z_s$ in Eq.~\eqref{eq:Zs} models the response of the --solely-- bulky sample under test. Whenever the e.m. penetration depth ($\delta_n$ in the normal state, $\lambda$ in the Meissner state) is in the order of the sample's thickness $d$, one should compute either $Z_s$ from the effective surface impedance~\cite{Jackson1975,Pompeo2025} or the finite-thickness-induced uncertainties to the measurement~\cite{Pompeo2025}. This is especially relevant in the normal state, when the e.m. penetration depth is the largest possible and is given by~\cite{Jackson1975}:
\begin{equation}
    \delta_n=[\rho_n/(\pi\nu\mu_0)]^{1/2}\textrm{;}
    \label{eq:deltan}
\end{equation}
or deep in the mixed state where the (complex) penetration depth~\cite{CoffeyClem1991}
\begin{equation}
    \tilde{\lambda}\approx[\lambda^2-\textrm{i}\tilde{\delta}_{vm}^2/2)]^{1/2}
    \label{eq:lambdatilde}
\end{equation}
is increased, starting from the Meissner value, by vortex motion.
\section{Measurement methods}
\label{sec:method}
A dielectric-loaded resonator (DR) is used to perform accurate measurements of the samples' $Z_s$, as described extensively elsewhere~\cite{Pompeo2014,Alimenti2019,Pompeo2021,Torokhtii2021}. In the inset of Fig.~\ref{fig:DR_T-sw}, the montage of the DCMS sample within the DR is shown. Both samples are measured using the same dielectric TiO$_2$ (rutile; ${\varepsilon_\parallel\approx250}$, ${\varepsilon_\perp\approx110}$ at ${T\lesssim20\,\textrm{K}}$~\cite{Breeze2016,Pompeo2014,Klein1995}), which yields similar spectra.
The DR is coupled through two small loops that terminate each K-type coaxial cable protruding the top cover (removed and not visible in the inset of Fig.~\ref{fig:DR_T-sw}), and connected to each port of a Vector Network Analyzer at the opposite ends. During measurements, the DR hangs fixed within the bore of a cryo-magnet capable of supplying ${\mu_0H\leq12\,\textrm{T}}$, meanwhile the temperature is controlled acting on heaters and fluxing He gas. Both field sweeps (named $H$-sw) at a fixed reduced temperature ${T/T_c:=1/3}$ (${T\approx6\,\rm{K}}$) and temperature sweeps (named $T$-sw) up to ${T \gtrsim T_c}$ at zero field (ZF) are performed. The spectrum at about the fundamental TE$_{011}$ (the electric field is orthogonal to the axis of the cylindrical dielectric, thus $\varepsilon\equiv\varepsilon_\perp$) is continuously tracked during $T$/$H$-sw. The measured resonant frequencies of the VTD and DCMS result ${\nu_0\approx8.50\,\rm{GHz}}$ and ${\nu_{0}\approx8.25\,\rm{GHz}}$, respectively (such difference comes from the use of the spacer shown in the inset of Fig.~\ref{fig:DR_T-sw}, which served to get similar coupling factors between samples despite their different shape). The scattering parameters $S_{ij}$, ${i,j\in\{1,2\}}$, are saved at each $\mu_0 H$ (in $H$-sw) or $T$ (in $T$-sw), and then fitted: $\{S_{ii}\}$ are elaborated following~\cite{Leong2002} to obtain the coupling factors $\{\beta_i\}$; whereas the Lorentzian-like resonance curves $S_{ij}$, ${i\neq j}$, are fitted as described in~\cite{Torokhtii2021}, yielding the loaded quality factor ${Q_L=Q_0[1+\beta_1+\beta_2]^{-1}}$ and the resonance frequency $\nu_0$.
In Fig.~\ref{fig:Q0Df0}, the unloaded quality factors $Q_0(H)$ and frequency shifts ${\Delta\nu_0(H):=\nu_0(H)-\textrm{max}[\nu_0(H)]}$ at ${T\approx6\,\textrm{K}}$ concerning the VTD and DCMS sample measurements are shown ($T$-sw is performed analogously). At ZF, it is directly noticeable that both samples yield $Q_0$ in the same order of magnitude. The $\textrm{max}[Q_0(H)]$ is roughly saturated by $R_s$ of the resonator's upper copper plate and the arrangements inside the DR. However, this is little relevant to the analysis since $\Delta Z_s$ at high magnetic field is the primary interest.
So as to obtain $\Delta Z_s$, the resonating mode in the cavity as it is assembled (see the inset in Fig.~\ref{fig:DR_T-sw}) is e.m. simulated using the eigenmode solver of \textit{CST Studio {Suite}~\textregistered}~2025 SP2~\cite{CST}, leading to the sample geometrical factor: ${G_{s}\approx204\,\Omega}$ for the VTD sample, and ${G_{s}\approx218\,\Omega}$ for the DCMS sample. The simulated frequencies are crosschecked against the measured ones, showing excellent agreement and thus validating the $\Delta Z_s$ calculations thereby. To remove the background, differential measurements are performed~\cite{Chen2004}:
\begin{itemize}
    \item On one hand, in $T$-sw, as $\varepsilon_\perp$ shows a strong dependence on $T$, though its loss tangent is approximately constant at $T\lesssim20\,\textrm{K}$~\cite{Pompeo2014,Klein1995}, it is just computed 
    \begin{equation}
        \begin{split}
            \hspace{-0.225in}\Delta R_{s}(T)=G_{s}\left(1/Q_{0}(T)-1/\textrm{max}[Q_{0}(T)]\right)\textrm{.}
        \end{split}
        \label{eq:Rs_T-sw}
    \end{equation}
    \item On the other hand, in $H$-sw, as there is no magnetic material but the sample, it follows that
    \begin{equation}
        \begin{split}
            \Delta Z_{s}(H) = & G_{s}\left(1/Q_{0}(H)-1/\textrm{max}[Q_{0}(H)]\right)- \\
            & - 2{\rm{i}}G_{s}\left(\nu_{0}(H)/\textrm{max}[\nu_{0}(H)]-1\right)\textrm{.}
        \end{split}
        \label{eq:Zs_H-sw}
    \end{equation}
\end{itemize} The results and discussions thereof are presented below.

\section{Results}
\label{sec:results}
\subsection{Temperature sweeps}
\label{sec:T-sw}
In Fig.~\ref{fig:DR_T-sw}, $\Delta R_s(T)$ of both samples in zero dc magnetic field is represented as computed by means of Eq.~\ref{eq:Rs_T-sw}. By direct inspection of the figure, the differences between the coatings can be appreciated: (i) the VTD sample presents ${T_{c}\approx18.0\,\rm{K}}$, whereas the DCMS sample shows ${T_{c}\approx17.2\,\rm{K}}$, consistent with the results presented in~\cite{Fonnesu2026}. It is well known~\cite{Godeke2006,Mentik2016} that $T_c$ of Nb$_3$Sn is correlated with the intermetallic composition and lattice parameter, as well as $\rho_n$ is affected by disorder and the presence of out-of-stoichiometry phases; (ii) the latter being in agreement with the normal state resistivity, which is obtained --exclusively in the bulk regime-- by means of Eq.~\eqref{eq:Zs}: ${\rho_n\propto R_s^2(T>T_c)\simeq\Delta R_s^2(T>T_c)}$ (the latter equivalence holds since the measured residual resistance of both samples~\cite{Posen2015,Fonnesu2026} is negligible in the order of magnitude of the normal state $R_s$). 
While in the normal state the DCMS sample is thick enough to be evaluated as bulk, the VTD sample is roughly as thick as the field penetration depth ${\delta_n\approx1.8\,\upmu\textrm{m}}$ (from Eq.~\eqref{eq:deltan}, in which ${\rho_n=R_s^2(T>T_c)/(\pi\nu\mu_0)\approx11\,\upmu\Omega\textrm{cm}}$ as for Eq.~\eqref{eq:Zs}), making Eq.~\eqref{eq:Zs} not valid. Although this matter deserves a separate exhaustive analysis, in this work it is plausibly assumed that $\rho_n$ of the VTD sample is not higher than that of the DCMS sample (this is indeed corroborated in the considerations made on ${\rho_{\rm{ff}}\propto\rho_n}$ and the extrapolation from field sweeps as described in Section~\ref{sec:H-sw}); and (iii) the superconducting transition of the DCMS is noticeably wider than that of the VTD sample, which again points to a higher purity (single phase) of the latter with respect to the former. However, all these signs present in $\Delta R_s(T)$ cannot be taken as definitive for the performance at high magnetic field -- even though $\rho_n$ sets the scale of $\rho_{\textrm{ff}}$ --because the inhomogeneities in the Ginzburg-Landau parameter $\Delta\kappa$ and defects are important sources of pinning~\cite{Dew-Hughes1974} that eventually increase $\nu_p$. Thus, $\Delta Z_s(H)$ is discussed below.
\begin{figure}[!t]
\centering
\includegraphics[width=0.95\columnwidth]{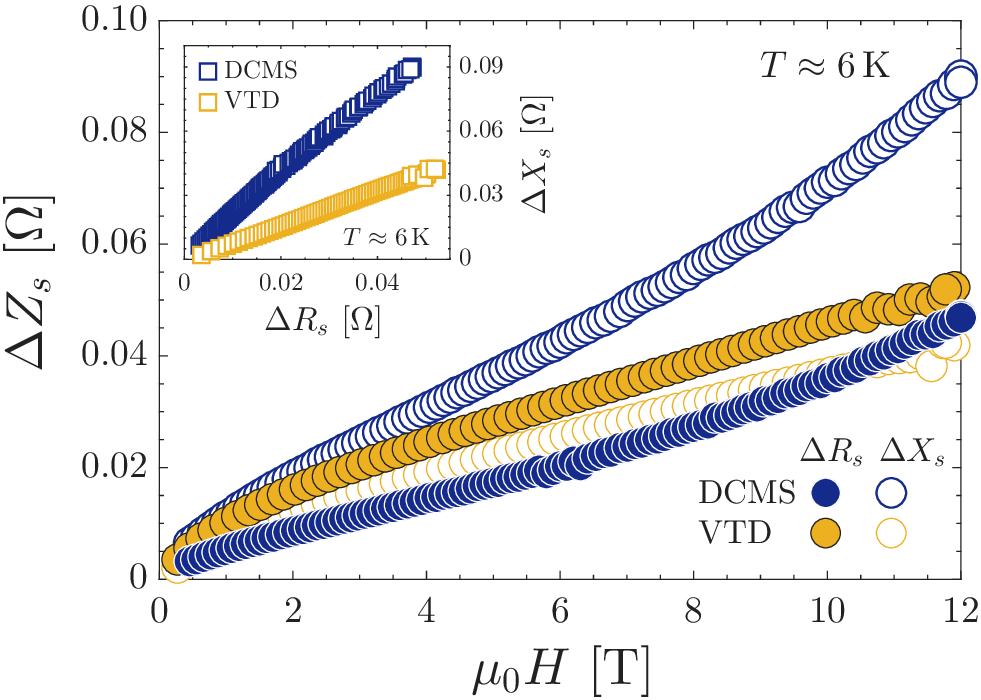}
\caption{Surface impedance variation of the VTD and DCMS sample as a function of the applied magnetic field ($\Delta Z_s(H)$), at $T\approx\,6\,\textrm{K}$. In the inset, the complex plane associated to $\Delta Z_s$ is represented to outline the differences between the samples' response (see discussion thereof in Section~\ref{sec:H-sw}).}
\label{fig:DZ_H-sw}
\end{figure}
\subsection{Field sweeps}
\label{sec:H-sw}
In Fig.~\ref{fig:DZ_H-sw}, $\Delta Z_s(H)$ at ${T\approx6\,\textrm{K}}$ for the DCMS and VTD samples is represented as computed by means of Eq.~\ref{eq:Zs_H-sw}. At this low temperature, and in low-to-moderate dc fields, it is expected that (a) the bulk expression{, Eq.~(1)),} largely holds {(the e.m. field does not penetrate completely the film)} and (b) the field-dependence of the two-fluid conductivity is negligible, so that $\Delta Z_s(H)$ roughly represents the excess of impedance from the Meissner state caused by vortex motion.

In particular, for the thinner VTD sample a check about the actual e.m. field penetration can be done resorting to Eq. \eqref{eq:lambdatilde}. By taking ${\rho_{\textrm{ff}}=\rho_n \mu_0H/B_{c2}}$, being $\mu_0H/B_{c2}\lesssim0.5$ at $T=6\;$K, and neglecting pinning, one obtains $\tilde{\delta}_{vm}=\delta_n/\sqrt{2}$. Thus, the complex propagation factor $\exp(-d/\tilde{\lambda})$ can be computed, obtaining an e.m. field attenuation to $\approx20\%$ when the substrate is reached, which allows neglecting the substrate contribution as a first approximation.

Both coatings roughly show (i) the same $\Delta R_{s}$, but (ii) significantly different variation of surface reactance $\Delta X_{s}$. This is appreciable also in the slope of the respective curves in the complex plane $\{\Delta R_s , \Delta X_s\}$ shown in the inset of Fig.~\ref{fig:DZ_H-sw}. This simple comparison between samples allows to assert that, based on Eq.~\eqref{eq:rho} and plausibly assuming that both coatings shown comparable $\lambda$, the mechanisms that yield such comparable $\Delta R_{s}$ but different $\Delta X_{s}$ originate from different pinning regimes. Actually, there is no way that both features take place simultaneously at comparable $\tilde{\rho}_{vm}$ and $\nu_0$. In the following, the samples are analyzed separately to figure out an estimation of pinning parameters.

First, by looking at $\Delta Z_s(H)$ of the VTD sample, it is noticeable that the curves evolve into ${\Delta X_s/\Delta R_s\approx1}$. At the same time, the dynamical range of $\Delta X_s(H)$ is similar to the maximum variation of ${\Delta R_s(T)\approx\Delta X_s(T)}$ in $T$-sw, so that it is reasonable to assume that ${\textrm{max}[\Delta X_s(H)] \gg X_s(H=0)}$. This means that, asymptotically when ${\mu_0H \to B_{c2}}$, $\Delta Z_s(H) \sim Z_s(H)$. Moreover, in the same limit, it is expected that $|\tilde{\rho}_{vm}|\gg\lambda^22\pi\nu\mu_0$ in Eq.~\ref{eq:rho} (as ${\rho_{\rm{ff}}\to\rho_n}$, and ${\nu_p\to0}$). As a result, the fact that $\Delta X_s/\Delta R_s\approx1$ already from moderate fields is a nice indication that ${\nu_p<\nu_0\approx8\;}$GHz (see a similar analysis done in~\cite{Tsuchiya2000}), with ${\nu_p}$ reducing to low values (below 1 GHz, in comparison with $\nu_0$) already at low fields. In fact, it is the presumably low value of $\rho_n$ that yields the limited $\Delta Z_s(H)$ observed. It is worth stressing that ${\Delta X_s \simeq \Delta R_s}$ at low fields is possible when $\tilde{\rho}_{vm}$ is real, that is $\nu_p \ll \nu_0$. Thus, in this particular case, the data point to low-to-negligible pinning.

Second, bearing in mind the considerations above when approaching $B_{c2}$, it is clear that the pinning regime of the DCMS sample, which exhibits ${\Delta X_s(H)>\Delta R_s(H)}$, is substantially different yielding a pinning frequency $\nu_p$ reasonably larger than $\nu_0\approx8\;$GHz. This is even more clear when comparing $\rho_n$, since the only way to limit $\Delta R_s(H)$ when $\rho_{\rm{ff}}$ is notably high is by enhanced pinning. Indeed, the possible presence of $\Delta\kappa$ and/or other defects could justify such considerable pinning strength.
\section{Conclusion}
\label{sec:conclusion}
A preliminary analysis on the magnetic-field-induced surface impedance variation of Nb$_3$Sn coatings grown via two different deposition techniques originally developed for accelerating cavities, VTD and DCMS, is presented with prospects of being used in haloscopes. By varying field, and temperature, fundamental considerations on the pinning properties and vortex-motion of the samples in the microwave range are deduced: the VTD sample presumably performs in the free-flow regime, already from few tesla; whereas the response of the DCMS sample is in a different pinning regime, at least up to several tesla. While the absolute level of dissipation are similar for VTD and DCMS samples, they originate from very different origins, namely different normal state resistivities and pinning strength.

Future work will be focused on analyzing quantitatively the obtained results, looking for the specific possible origins of pinning, and identifying the underlying mechanisms that determine the flux flow resistivity. In order to reach this goal consistently, the present analyses will be extended by exploring the response at different frequencies and taking carefully and analytically into account the possible effects of partial e.m. penetration to the substrate at high fields. In addition, the impact of the layer thickness and its possible influence on the morphology of grains and boundaries in between will be tackled by testing diverse depositions and substrates/buffer layers.

Ultimately, the accurate determination of the vortex motion parameters will allow to estimate $Z_s$, and then the main figures of merit of different haloscope cavities as they were coated with those films.
\end{document}